\documentclass[twocolumn,prl,preprintnumbers,tightenlines,nofootinbib,groupedaddress]{revtex4-1}
\usepackage{graphicx}
\usepackage{dcolumn}
\usepackage{bm}
\usepackage{amsmath}
\usepackage{amsfonts} 
\usepackage{latexsym}
\usepackage{bbm}
\usepackage{color}
\usepackage{amssymb}
\usepackage{amsthm}
\usepackage{epsf}
\usepackage{epsfig}
\usepackage{caption3}
\usepackage{appendix}
\usepackage{cases}
\usepackage{subfigure}
\usepackage{multirow}
\usepackage{url}
\usepackage{hyperref}
\makeatletter

\newcommand{\Rmnum}[1]{\expandafter\@slowromancap\romannumeral #1@}
\makeatother

\begin{document}

\title{Fast radio bursts from axion stars moving through pulsar magnetospheres}

\author{James H. Buckley}%
\email{buckley@wustl.edu}
\author{P. S. Bhupal Dev}%
\email{bdev@wustl.edu}
\author{Francesc Ferrer}%
\email{ferrer@wustl.edu}
\author{Fa Peng Huang}%
\email{fapeng.huang@wustl.edu (corresponding author)}

\affiliation{Department of Physics and McDonnell Center for the Space Sciences, Washington University, St. Louis, MO 63130, USA}

	
\begin{abstract}
We study the radio signals generated when an axion star enters  the
magnetosphere of a neutron star. As the axion star moves
through the resonant region where the plasma-induced photon mass becomes equal to the axion mass, the axions can efficiently convert into photons, giving rise to an intense, transient radio signal. 
We show that a dense axion star with a mass $\sim 10^{-13}M_{\odot}$ composed of
$\sim 10\: \mu$eV axions can account for most of the mysterious 
fast radio bursts.
\end{abstract}
\maketitle

Weakly coupled pseudoscalar particles such as axions, that arise from a
solution to the strong CP-problem~\cite{Peccei:1977hh,*Weinberg:1977ma,*Wilczek:1977pj,*Kim:1979if,*Shifman:1979if,*Dine:1981rt,*Zhitnitsky:1980tq}, or more
generic axion-like particles (ALPs) predicted
by string theory~\cite{Svrcek:2006yi,*Arvanitaki:2009fg,*Cicoli:2012sz},
are promising dark matter (DM) candidates and may contribute significantly to
the energy density of the
Universe~\cite{Preskill:1982cy,*Abbott:1982af,*Dine:1982ah}.
In recent years,  an increased interest on axion DM has bolstered a broad
experimental program~\cite{Irastorza:2018dyq},
often based on the Primakoff process~\cite{Pirmakoff:1951pj}, 
whereby axions transform into photons in external magnetic fields and vice versa.

Low mass  axions or ALPs that contribute
appreciably to the DM must have extremely high occupation numbers,
and can be modeled by a classical field condensate. 
Such large number density in astrophysical
environments 
enables to probe their existence indirectly
through the detection of low energy photons. 
For $\mu$eV-scale axions
consistent with the observed DM density,  the emitted photons
have frequencies in the range probed by radio telescopes.
Along these lines,
signals resulting from the axion decay to two
photons~\cite{Caputo:2018ljp,*Caputo:2018vmy}, or from resonant axion-photon
conversion~\cite{Pshirkov:2007st,Huang:2018lxq,Hook:2018iia} have been recently
explored.

If the Peccei-Quinn (PQ) symmetry~\cite{Peccei:1977hh} is broken after inflation, the axionic DM
distribution is expected to be highly inhomogeneous, leading to  the formation
of axion miniclusters as soon as the Universe enters the matter-domination
regime~\cite{Hogan:1988mp,*Kolb:1993zz,*Kolb:1993hw}, which in turn may lead to  the formation
of dense boson stars~\cite{Kaup:1968zz,*Ruffini:1969qy} that
could make part of the DM~\cite{Eggemeier:2019khm}. Such boson stars 
are called axion stars, when the kinetic
pressure is balanced by self-gravity, or axitons, when stabilized by
self-interactions (see Ref.~\cite{Braaten:2019knj} for a recent review).
Gravitational microlensing
could potentially constrain the fraction of DM in collapsed
structures~\cite{Fairbairn:2017dmf}, but typical axion star signals fall in the
femtolensing regime 
which is not robustly constrained~\cite{Katz:2018zrn}. Although their
presence may be unveiled in future by observations of highly magnified
stars~\cite{Dai:2019lud}, it is important to look for other 
experimental probes.

Such dense clumps of axion DM can lead to enhanced radio signals, which might explain the mysterious observation of Fast Radio Bursts (FRBs)~\cite{Lorimer:2007qn, Thornton:2013iua}. For instance,
the oscillating axion configuration 
of a dilute axion star hitting the
atmosphere of a neutron star 
could induce dipolar radiation of the
dense electrons in the atmosphere~\cite{Iwazaki:2014wka} or neutrons
in the interior~\cite{Raby:2016deh}, and generate a powerful radio 
signal.
However, as noted in Ref.~\cite{Pshirkov:2016bjr}, the dilute star
will be tidally disrupted well before reaching the surface of the neutron star.
Moreover, the plasma mass of a photon radiated at the surface of the neutron star 
is much larger than the frequency of the dipole
radiation. 
Hence, medium effects would greatly suppress the signal. 

Even the optimistic scenario of a dense axion star directly hitting the
NS surface would lead to, at most, a $\mu$Jy radio 
signal~\cite{Bai:2017feq},
whereas FRBs range from $\mathcal{O}$(0.1) to $\mathcal{O}$(100)~Jy.
Their large dispersion measure suggests that the FRBs are  of extragalactic
origin, 
$0.1\lesssim z \lesssim 2.2$, which
means that the total energy released 
is about $\mathcal{O}(10^{38}-10^{40})$~erg, and their observed
millisecond duration requires that the radiated power reaches
$10^{41}$--$10^{43}~{\rm erg\cdot s}^{-1}$.
Although their origin and physical nature are still
obscure~\cite{Popov:2018hkz,*Ye:2017lqn,*Deng:2018wmy,*Sun:2020gem},
the fact that the energy released by FRBs is about 
$10^{-13}M_{\odot}$,
 which is the typical axion star mass,
and that their frequency (several hundred MHz to several GHz)
coincides with that expected from $\mu$eV axion particles,
motivates us to
further explore whether the axion-FRB connection can be made viable in 
an neutron star environment and tested with future data.\footnote{See Refs.~\cite{Tkachev:2014dpa,Rosa:2017ury} for alternative
proposals not involving neutron star.}
\begin{figure}
  \begin{center}
    \includegraphics[width=0.48\textwidth]{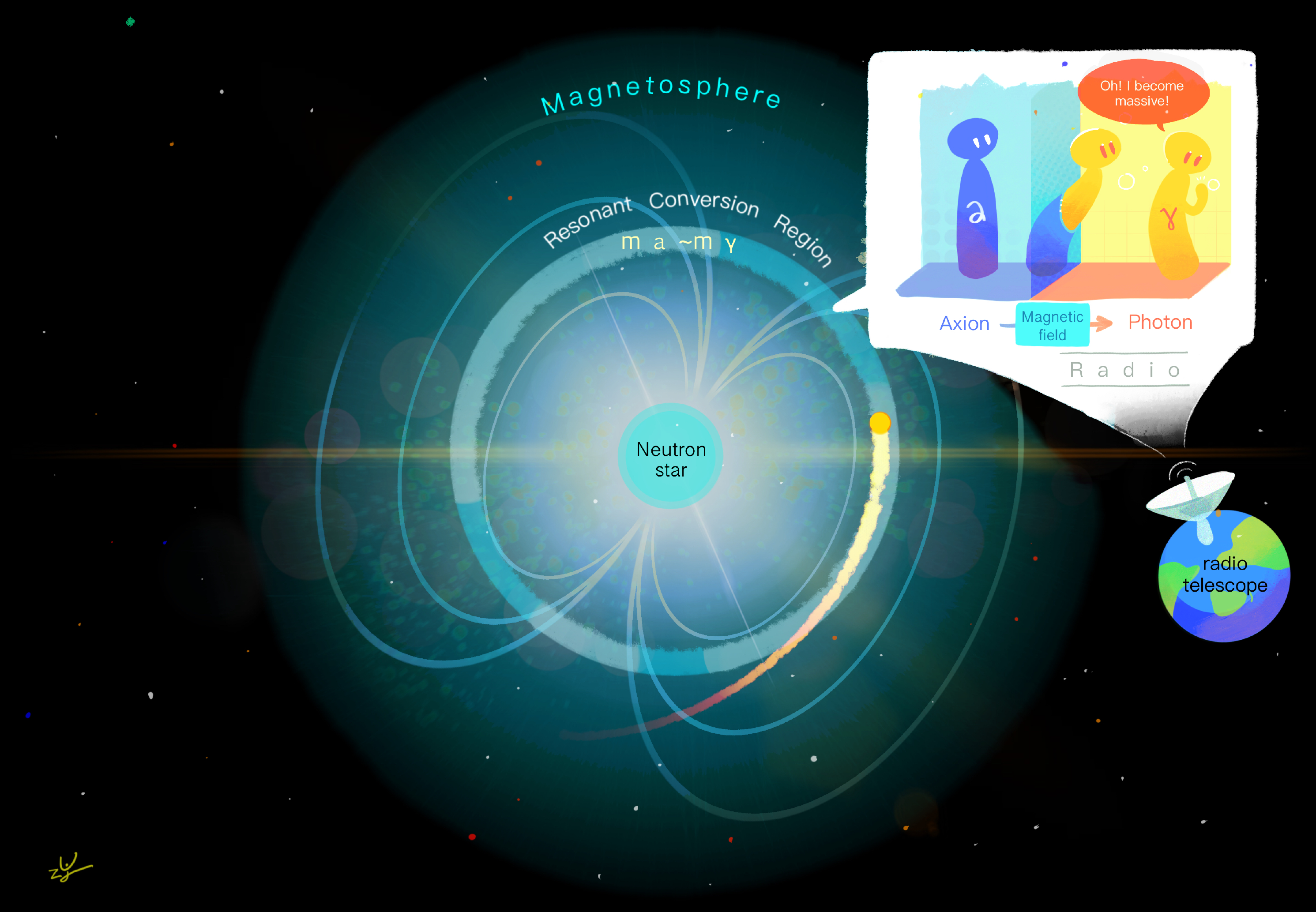}
\end{center}
\caption{Schematic diagram of the proposed FRB
 signal from dense
	axion star. When a dense axion star passes through the resonant
	conversion region in the magnetosphere of  neutron star (where the
	effective photon mass equals the axion mass), powerful transient
	radio signals can be produced in the strong external magnetic field
	through the Primakoff process.}\label{pr}
\end{figure}

In this paper, we propose a new explanation for 
FRBs based on the
resonant axion-photon conversion that takes place when a
dense axion star passes through the {\em resonant region} in the
magnetosphere of an neutron star, as shown in
Fig.~\ref{pr}.
We will mainly focus on non-repeating FRBs in this work, since repeating FRBs may
correspond to a different source class~\cite{Katz:2019rwy}.  
So far, more than 60 non-repeating FRBs have been
observed~\cite{Petroff:2016tcr,frbcat} mainly by
Parkes, ASKAP,  and UTMOST.
Our explanation of the non-repeating FRB signals roughly from 800~MHz to 1.4~GHz
involves dense stars made of axions with mass about 10~$\mu$eV.
%

The properties of an axion star depend on its mass $M_a$, axion mass
$m_a$ and decay constant $f_a$. Dilute axion stars
have a radius~\cite{Chavanis:2011zm}
\begin{align}
	R_a^{\text{dilute}} & \sim \frac{1}{G_NM_a m_a^2} \nonumber \\
	& \cong
	(270~\text{ km}) \left(\frac{10~\mu\text{eV}}{m_a}\right)^2
	\left(\frac{10^{-12} M_{\odot}}{M_a}\right) .
\end{align}
Hence, the typical radius of a
dilute axion star is $\mathcal{O}(100)$ km for the star mass
range $M_a \sim 10^{-14}-10^{-12} M_{\odot}$.
A  dense star branch was first proposed
in Ref.~\cite{Braaten:2015eeu}.
Nevertheless, it was pointed out in Ref.~\cite{Visinelli:2017ooc} that  axion 
field values reach $\gtrsim {\cal O} (1)$ 
in the core, thus making the axions relativistic and rendering the analysis
in Ref.~\cite{Braaten:2015eeu} inconsistent (see
also Refs.~\cite{Chavanis:2017loo,Schiappacasse:2017ham,Eby:2019ntd}).
Since gravity is negligible inside such dense stars, their profiles can instead
be found as solutions of a Sine-Gordon type equation leading
to their natural identification 
with oscillons. In contrast
to the natural expectation that localized, finite energy configurations of
the axion field decay within $\tau \sim 1/m_a$, oscillons can last
$\cal{O}$ (100-1000)
oscillations~\cite{Bogolyubsky:1976yu,*Gleiser:1993pt,*Gleiser:1999tj,*Fodor:2006zs,*Salmi:2012ta}, before disappearing into a burst of relativistic
axions~\cite{Kolb:1993hw}.
For a QCD axion, these timescales
still fall short of being of cosmological relevance. Nevertheless, flatter
potentials at large field values in well motivated ALP models have been shown 
to feature much longer-lived oscillons,
$\tau > \left(10^{8-9}\right)/m_a$, and for plateau-like potentials only lower
bounds on their lifetime are known~\cite{Olle:2019kbo}. 
Stable dense profiles are also possible when
$f_a \gtrsim 0.1 M_{\text{Pl}}$~\cite{Helfer:2016ljl}. On the other hand,
axion stars could have been created much after matter domination
via parametric amplification of axion fluctuations even if the PQ symmetry is 
broken before inflation~\cite{Olle:2019kbo,Arvanitaki:2019rax}. Given that 
oscillons are attractor solutions, it cannot be excluded that dense axion 
configurations are being generated and are present in astrophysical settings 
such as pulsars~\cite{Garbrecht:2018akc}. 
In this work, we assume that dense axion stars with 
a mass around
$10^{-13}M_{\odot}$ can survive to the present, and have a chance to 
encounter an neutron star.  
For dense axion stars, the radius can be  approximated as~\cite{Braaten:2015eeu}
\begin{align}
	R_a^{\text{dense}} & \sim  (0.47~\text{m}) \sqrt{g_{a\gamma\gamma}\times 10^{13}~\text{GeV} 
	\frac{10~\mu\text{eV}}{m_a}} \nonumber \\
	& \quad \times 
	\left(\frac{M_a}{10^{-13}M_{\odot}}\right)^{0.3}, 
\end{align}
with $g_{a\gamma\gamma}$ being the axion-photon coupling.  

Tidal effects become important when the distance of the axion star to the neutron star 
approaches the Roche limit:
\begin{equation}
	r_t=R_a \left(  \frac{2 M_{\rm NS}}{M_a} \right)^{1/3},
\end{equation}
where $M_{\rm NS}$ is the neutron star mass.
A  gravitationally bound object approaching a star closer than this radius
will be disrupted by tidal effects~\cite{Tkachev:2014dpa,Pshirkov:2016bjr}.
A 100 km dilute axion star will be destroyed at $r_t \sim 10^6$ km,
long before it enters the magnetosphere. Tidal disruption may quickly 
rip the dilute axion star apart, producing
a stream of axion debris that would then be swallowed by the neutron star.
It is conceivable that the subsequent interaction of the tidal debris with
the neutron star could lead to multiple radio signals, similar to the repeating
FRBs~\cite{Spitler:2016dmz, Andersen:2019yex, Amiri:2020gno}, and this possibility deserves further investigation.

For a dense axion star, however, the radius is smaller than a meter
and the Roche limit is about 10 km.
Thus, a dense axion star can reach the resonant conversion region. 
Tidal forces will certainly stretch the axion star
in the radial direction and compress it in the transverse direction.
Since the resonant conversion region is located over a hundred Schwarzschild
radii from the neutron star, we can use Newtonian gravity to estimate
the tidal deformation:
\begin{equation}
\frac{\delta R_{a}}{R_{a}}=\frac{9 M_{\rm NS}}{8\pi \rho_{\rm AS} r^3}\, ,
\end{equation}
where $\rho_{\rm AS}$ is the axion star density and $r$ is its distance from the neutron star. For a typical dense axion star with $M_a \sim 10^{-13}M_{\odot}$, 
the tidal deformation is negligible, $\mathcal{O}(10^{-3})$.


When a dense axion star enters the magnetosphere, axions convert to 
radio signals through 
the 
axion-photon interaction term
\begin{eqnarray}
\mathcal{L}=-\frac{g_{a\gamma\gamma}}{4} a F^{\mu\nu} \tilde{F}_{\mu\nu}
=g_{a\gamma\gamma}  a \vec{E} \cdot \vec{B}  \,\,,
\end{eqnarray}
where $a$ is the axion field, $F^{\mu\nu}$ the electromagnetic field
strength, and $\tilde{F}{}^{\mu\nu}$ its dual.
For 
$m_a\sim 10~\mu$eV, the coupling is constrained to be 
$g_{a\gamma\gamma}\leq 10^{-13}~\rm GeV^{-1}$~\cite{Hagmann:1990tj,Tanabashi:2018oca}.
Neutron star magnetospheres, featuring
the strongest magnetic fields known in the Universe, provide one of the best environments for axion-photon conversion. Due to the extremely small coupling
$g_{a\gamma\gamma}$, however,
the conversion probability is generally expected to be small. 
On the other hand, the conversion can be significantly enhanced in the
resonant conversion region of the magnetosphere.
Indeed, the photon acquires a mass due to the plasma effects in the
magnetosphere~\cite{Raffelt:1996wa} :
\begin{equation}\label{mp}
	m_{\gamma}(r)\simeq \omega_{p}=\sqrt{\frac{e^2 n_e}{m_e}}
	=\sqrt{\frac{n_e}{7.3 \times 10^{8} \rm~cm^{-3}}} ~\mu\text{eV ,}
\end{equation}
where $n_e(r)$ is the local electron density at a distance $r$ from the 
neutron star center.
For simplicity, we use the Goldreich-Julian 
distribution~\cite{Goldreich:1969sb}:
\begin{equation}\label{ne}
	n_e(r)=7 \times 10^{-2} \frac{1 \text{ s}}{P}
	\frac{B(r)}{1 \text{ G}}\text{ cm}{}^{-3},
\end{equation}
where $P$ is the rotation period of the neutron star. For the magnetic field $B(r)$, we take the dipole approximation: 
\begin{equation}\label{br}
	B(r)=B_0 \left( \frac{r_{\text{NS}}}{r}\right)^3 \, , 
\end{equation}
with $B_0$ being the magnetic field strength at the neutron star surface, 
which can reach $10^{15}$~G for a magnetar~\cite{Weber:1999qn}.
The scale of magnetosphere is  $\mathcal{O}(100)$
$r_{\text{NS}}\sim 1000$~km.

Note that
	QED vacuum polarization effects can also contribute to the photon
mass~\cite{Raffelt:1987im} $m_{\gamma}^2=\omega_{p}^2-m_{\rm QED}^2$, with
$m_{\rm QED}^2=\frac{7e^2}{180\pi^2}\omega^2 \frac{ B^2}{B_{c}^2} $ and
$B_{c}=m_e^2/e\approx4.4 \times 10^{13}$\ G.
However, comparing the two contributions, 
\begin{equation}\label{cc}
\frac{\omega_{p}^2}{m_{\rm QED}^2} \sim  5\times 10^8 \left(\frac{\mu {\rm eV}}{\omega}\right)^2 \frac{10^{12} \text{ G}}{B}\frac{1 \text{ s}}{P},
\end{equation}
we see that the QED mass term becomes negligible
in our case with typical axion energy 
$\omega \sim 10 \ \mu $eV~\cite{Huang:2018lxq}.

In the resonant conversion region, the photon effectively has almost 
the same mass as the axion due to plasma effects:
\begin{equation}
	\label{eq:mm}
	\omega^2=k_a^2+m_a^2 \approx m_{\gamma}^2(r_c) , 
\end{equation}
where $\omega$ is the axion-photon oscillation frequency.
The mass degeneracy leads to maximal mixing and greatly enhances the conversion
probability. The critical radius $r_c$ for the resonant conversion region is
obtained by enforcing the maximal mixing condition Eq.~\eqref{eq:mm}:
\begin{equation}\label{rc}
	\left(\frac{r_{\rm NS}}{r_c}\right)^3 \sim \left(\frac{m_a}{\mu\text{eV}}
	\right)^2 \frac{10^{10}\text{ G}}{B_0} \frac{P}{1 \text{ s}}.
\end{equation}

When the dense axion star approaches $r_c$, resonant
axion-photon conversion can occur.
For most neutron star environments,  the resonant conversions are 
non-adiabatic~\cite{Huang:2018lxq}, with 
the conversion rate obtained as 
$P_{a\to \gamma} \approx 2\pi \beta$
with
\begin{equation}
	\beta=\left.\frac{\left( g_{a\gamma\gamma}\omega B_0 \right)^2/
	2\bar{k}}{\left| \text{d}\omega_{p}^2/\text{d}r \right|}
	\right|_{r=r_c}.
\end{equation}
Here, $\bar{k}\equiv \sqrt{\omega^2-(m_a^2+\omega_{p}^2)/2}$ is the axion 
momentum in the diagonalized basis of the mixing equations.
From Eqs.~\eqref{mp}-\eqref{rc},
we can derive
\begin{equation}
	\left. \frac{\text{d} \omega_{p}^2}{\text{d}r}\right|_{r=r_c}
	=\left. \frac{3 \omega_{p}^2}{r}\right|_{r=r_c}.
\end{equation}
We note that for typical parameters, close to the neutron star surface
$r< r_c$,  the effective photon mass is larger than the axion mass, and 
the emission of a photon is kinematically suppressed, impacting the
viability of the mechanisms proposed in 
Refs.~\cite{Iwazaki:2014wka,Bai:2017feq}.
%

As a dense axion star moves through the resonant region, the conversion power is $\dot{W} =  P_{a\to \gamma} \text{d} M_a/\text{d}t  $
with $\text{d}M_a/\text{d}t\sim\pi R_a^2\rho_A v_c$ and
$\rho_A=M_a/(4\pi R_a^3/3)$.
Thus, we obtain the power:
\begin{equation}
	\label{power}
	\dot{W} \sim  \left(\frac{M_{a}}{10^{-13}M_{\odot}}\right)  \left(10^7 \times P_{a\to \gamma}\right)\left(10^{44}~\text{GeV}\cdot
	\text{s}{}^{-1}\right) \, .
\end{equation}
For the benchmark values $B_0=10^{14}$ G, $m_a=10~\mu$eV,
$g_{a\gamma\gamma} = 10^{-13}~\text{GeV}{}^{-1}$, conversion in a typical
$1.4 M_\odot$ pulsar rotating with $P=0.1$~s occurs with
$P_{a\to \gamma} \approx 2 \times 10^{-5}$ in the resonant region.
Hence, a $10^{-13}M_{\odot}$ dense axion star can naturally account for
the typical output associated to FRBs,
$\dot{W} \sim 10^{44}~\rm GeV\cdot s^{-1}  $.
The trajectory of a dense axion star moving roughly parallel to the 
resonant region is schematically shown in Fig.~\ref{pr}.
Upon entering, the axion star moves in the resonant region for about 10 km (several milliseconds) 
until it leaves the region or evaporates. 

To compare with the current FRB data~\cite{Petroff:2016tcr,frbcat}, 
we use the convention:
\begin{equation}
\label{exp}
	\frac{E_{\text{FRB}}}{\text{J}}=
	\frac{F_{\text{obs}}}{\text{Jy}\cdot \text{ms}}
	\frac{\Delta B}{\text{Hz}} \left(\frac{d}{\text{m}}\right)^2
	\times 10^{-29} (1+z),
\end{equation}
where $E_{\text{FRB}}$ is the FRB energy ($10^{30}$ to $10^{33}$~J), 
$d$ is the source distance
(from several hundred Mpc to several Gpc), and $z$ is the redshift.
$\Delta B$ is chosen as the bandwidth of the
radio telescope in current experiments~\cite{Petroff:2016tcr,frbcat}.
The fluence $F_{\text{obs}}$ is the density flux $\mathcal{S}\sim \dot{W}/(4\pi d^2 \Delta B)$ integrated over time.

\begin{figure}[t!]
	\centering
	\includegraphics[scale=0.25]{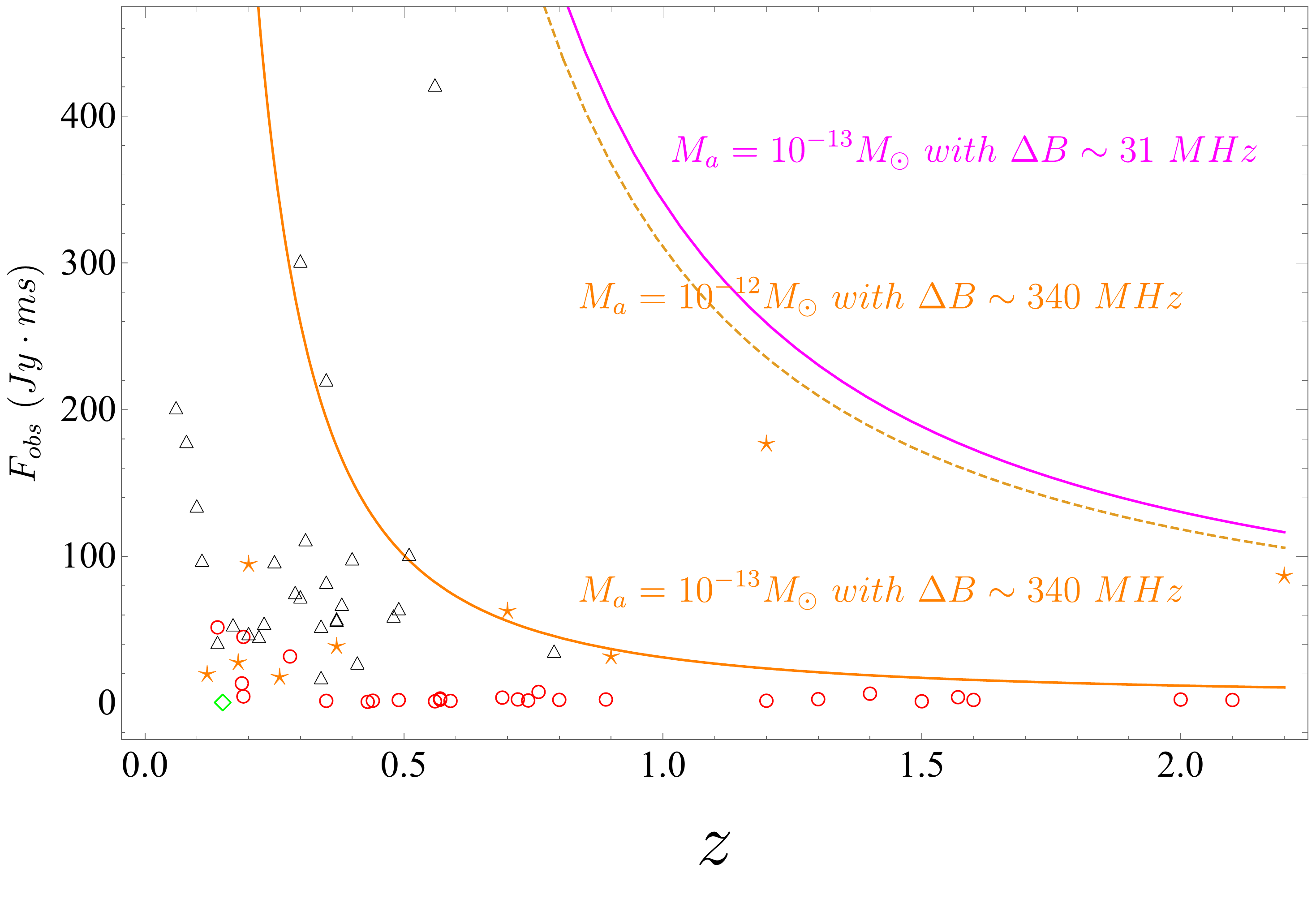}
	\caption{Upper limit on the fluence as a function of redshift $z$. The solid orange line
	depicts the upper limit for $M_a=10^{-13} M_{\odot}$ with bandwidth
	$\Delta B=340$~MHz. Similarly, the dashed orange and solid magenta lines are the upper limits for $(M_a,\Delta B)=(10^{-12} M_{\odot}, 340~{\rm MHz})$ and $(10^{-13} M_{\odot}, 31~{\rm MHz})$, respectively. The red circles, black triangles, green diamonds and orange stars show the 27 non-repeating FRBs observed by Parkes ($\Delta B\sim338.381$~MHz),
28 events from ASKAP ($\Delta B\sim336$~MHz), 1 event from Arecibo ($\Delta B\sim322.6$~MHz)
and 9 events from UTMOST ($\Delta B\sim31.25$~MHz), respectively~\cite{frbcat}.}
	\label{limit}
\end{figure}

For our benchmark values
we can naturally
explain most of the observed  FRBs
as shown in Fig.~\ref{limit}.
The orange line in Fig.~\ref{limit} depicts the upper limit for
$M_a=10^{-13} M_{\odot}$ with $\Delta B=340$~MHz, and
the events below this line can be accounted for.
The dashed orange line represents the upper limit for $M_a=10^{-12} M_{\odot}$
and the same bandwidth, while we used  $M_a=10^{-13} M_{\odot}$ and
$\Delta B=31$~MHz for the magenta line.
The red circles, black triangles, green diamonds and orange stars represent the
27 non-repeating FRBs observed by Parkes ($\Delta B\sim338.381$~MHz),
28 events from ASKAP ($\Delta B\sim336$~MHz), 1  event from Arecibo ($\Delta B\sim322.6$~MHz) and 9
events from UTMOST ($\Delta B\sim31.25$~MHz)~\cite{frbcat}, 
respectively.
Most events lie below the solid orange curve,
except a few events which can only be explained by a heavier axion
star, as shown by the dashed orange curve. For a smaller bandwidth $\Delta B=31$~MHz, even $M_a=10^{-13} M_{\odot}$ could explain all the events by this scenario, as shown by the magenta curve.\footnote{Here we only list the non-repeating FRBs with frequencies favored by the 10\:$\mu$eV axion.
We do not include other non-repeating FRBs with frequencies lower than 800 MHz, like the events from
CHIME and Pushchino~\cite{frbcat}, which may be better explained by a lighter axion or by other sources.}
Thus, a dense axion star with mass $\sim 10^{-13} M_{\odot}$,
consistent with theoretical expectations~\cite{Braaten:2019knj},
can explain the radiated energy of most of the observed FRB events. 
Those events in Fig.~\ref{limit} that do not saturate this limit 
can be due to an axion star with a different mass
or to the particular astrophysical environment. For instance, the
conversion probability is determined by the neutron star properties, 
like the magnetic field distribution and size of the neutron star, 
which could in principle broaden the spectrum of axion star masses 
from what is considered in Fig.~\ref{limit}.

An FRB emitted with a frequency $\nu_0=m_a/2\pi=2.42~\rm GHz$$(m_a/10~\mu$eV) in
the axion rest frame will be observed at a lower frequency by the time it reaches a radio telescope
on Earth mainly due to the cosmological redshift:
\begin{equation}
       \nu=\frac{\nu_0}{1+z} .
\end{equation}
Given the different cosmological redshifts measured for different FRB events,
for a 10 $\mu$eV axion, the observed frequency ranges from 700 MHz to $2.1$ GHz.
Thus, we can explain most of the observed FRB events which fall in this frequency range. 

We stress that our aim is to explain the broad features of FRBs,
but there are a number of complicated astrophysical effects that are likely 
 important in describing their details. 
The magnetosphere geometry (e.g., the position of gaps and the neutral sheet) 
has a significant impact on
the observed signals. Moreover, there are likely to be significant
feedback effects in the conversion region.  As the axion star moves through 
the field and plasma comprising the magnetosphere, it may exert radiation 
pressure on the surrounding plasma.
Other factors can also affect the variation of signal strengths and duration. 
For example, for a fixed axion mass, a larger
pulsar period  means a smaller $r_c$, and hence larger
$B(r_c)$ which leads to a larger conversion probability. 

The existing observational data on the degree of polarization of 
non-repeating FRBs are limited and inconclusive. Only 9 events
have polarimetric data available~\cite{Petroff:2019tty}, and the picture
that emerges is unclear: some events appear to be completely unpolarized, 
some show only circular polarization, some show only linear polarization, 
and others show both~\cite{Petroff:2019tty}. 
Several factors can influence the specific polarization expected in our 
scenario. 
First of all, the photons produced in the magnetosphere due to axion
conversion might have different polarizations depending on the local 
environment or viewing geometry.
Moreover, additional axion-photon conversions can take place during the 
propagation of the FRB pulse through the intergalactic magnetic field. Each
conversion is expected to generate some degree of circular 
polarization regardless
of the initial polarization at the source, depending on the properties of 
the cosmological magnetic field~\cite{Payez:2009vi,*Payez:2011sh, *Horns:2012pp,*Masaki:2017aea}. Given these uncertainties and the lack of sufficient 
polarimetric data, a detailed analysis of the polarization signal in our 
scenario is beyond the scope of the current paper, but it is certainly worth 
exploring in the future whether this could be used to distinguish our model 
from other explanations of the FRB events.

Furthermore, the axion star and the pulsar could potentially
form a binary system via e.g. three-body interactions. In this case,
the orbiting axion star could pass through the pulsar magnetosphere several 
times to produce repeating FRBs~\cite{Spitler:2016dmz, Andersen:2019yex, Amiri:2020gno}. A larger mass would be critical for the axion stars to survive multiple transits of the resonant conversion region. We leave the details of this mechanism for future work.

The smallest flux density that can be detected by a radio telescope
can be written as:
\begin{align}
	{\mathcal S}_{\rm min} \approx 0.09 \text{ Jy }\left(\frac{1~\rm MHz}{\Delta B}
	\right)^{1/2}
	\left(\frac{1 \text{ ms}}{t_{\text{obs}}}\right)^{1/2}
	\left(\frac{10^3 \text{m}{}^2/\text{K}}{A_{\text{eff}}/
	T_{\text{sys}}}\right),
\end{align}
where $t_{\text{obs}}$ is the observation time and $ A_{\text{eff}}/T_{\text{sys}}$ is the effective area to system temperature ratio.
For example, the SKA Phase 1~\cite{Bacon:2018dui} with 
$ A_{\text{eff}}/T_{\text{sys}}=2.7 \times 10^3 \text{m}^2/\text{K}$,  assuming $\Delta B=1$~MeV and $t_{\text{obs}}=40$~ms, 
can detect a radio signal if $\mathcal{S}>{\mathcal S}_{\rm min}\sim 5\times 10^{-3}$~Jy within the frequency range  0.45 to 13~GHz.
The sensitivity is expected to increase by more than an order of magnitude in
Phase 2 of SKA, which will enhance its ability to detect even weaker FRBs.

The event rate can be estimated as
\begin{align}
N/{\text{year}} \approx \sigma v_0  n_{\text{AS}} n_{\text{NS}} f_{\text{NS}}
V \, ,
\end{align}
 where 
$\sigma=\pi b^2=\pi r_c^2 v_c^2/v_0^2 (1-2 G_N M_{\text{NS}}/r_c)^{-1}$ is the scattering cross section for the axion star with a virial velocity $v_0$ approaching the neutron star with an impact parameter $b$.
The number of axion stars
is given by
$n_{\text{AS}}=\kappa_{\text{AS}} \rho_{\text{DM}}/M_a  \approx
\kappa_{\text{AS}} \times 10^{11} \text{ pc}{}^{-3}$, with the galactic
DM density $\rho_{\text{DM}}\sim0.3 ~\rm GeV\cdot \text{cm}{}^{-3}$, while
$\kappa_{\text{AS}}$ is the fraction of the total DM density in dense
axion stars, and $f_{\text{NS}}$ represents the ratio of neutron stars with
magnetic fields larger than $10^{13}$~G. We thus have
$N/\text{year}=10^{-2}\kappa_{\text{AS}}  f_{\rm NS}$ in our galaxy.
The event rate per day in the Universe
is $ 10^{13} \kappa_{\text{AS}} f_{\text{NS}}/365\sim 1000$, if we
take  conservative  values of $\kappa_{\text{AS}}=10^{-2}$~\cite{Fairbairn:2017dmf,Katz:2018zrn} and $f_{\text{NS}}=10^{-5}$~\cite{Beniamini:2019bga} .
This scenario satisfies the condition that
the events should be sufficiently rare to ensure that the Galactic plane does not
dominate the spatial distribution of observed events~\cite{Katz:2018xiu}.

In conclusion, we have proposed a new explanation for the origin of FRBs, based
on the axion-to-photon conversion that ensues when a dense axion
star moves through the resonant region in the pulsar magnetosphere.
The observed FRB energy output is naturally obtained for axion stars with masses
around $10^{-13}~M_{\odot}$ if the axion-photon conversion proceeds through the
non-adiabatic resonant regime. Most of the observed frequencies for
non-repeating FRBs can be accommodated with a 10~$\mu$eV axion mass.


In the future, the unprecedented sensitivity of SKA and other
radio telescopes may unravel the spectral properties of FRBs.
The many observed events in the 0.7 to 2.1 GHz range
correspond to the same intrinsic peak frequency at the emission time,
which could provide further support for this scenario.
Together with laboratory measurements from axion haloscopes and weak 
radio signals from diffuse axion DM, SKA is expected to observe many more FRBs,
and might allow to pin down the correlation between FRBs,
axions and DM.

\acknowledgments
We are grateful to Shmuel Nussinov for enlightening discussions and critical comments, Raymond Co, Jonathan Katz, Oriol Pujol\`as and Yicong Sui for useful discussions, Yurong Zhao for the manga. The work of  JB, BD and FF is supported
in part by the U.S. Department of Energy under Grant No. DE-SC0017987.
FPH is supported by the McDonnell Center for the Space Sciences.
\bibliographystyle{apsrev4-1}
\bibliography{axfrb}
\end{document}